\documentclass[preprint,prx,preprintnumbers,amsmath,amssymb,longbibliography]{revtex4-2}

\usepackage{times}
\usepackage{graphicx}
\usepackage{amsmath}
\usepackage{mathtools}
\usepackage{float}

\newcommand{\SNprojections}{Suppl. Note 1}
\newcommand{\SNlieb}{Suppl. Note 2}
\newcommand{\SNDCfit}{Suppl. Note 3}
\newcommand{\SNoxides}{Suppl. Note 4}
\newcommand{\SNbackscattering}{Suppl. Note 5}
\newcommand{\SNhybrid}{Suppl. Note 6}

\begin{document} 

\title{Halide Perovskites as Spin-1 Dirac Materials} 

\author{Dmitry Marchenko,$^{1}$ Maryam Sajedi,$^{1}$ Maxim Krivenkov,$^{1}$ Saleem Ayaz Khan,$^{2}$ Andrei Varykhalov,$^{1}$ Alexander Fedorov,$^{1,3,4}$ Jaime S\'anchez-Barriga,$^{1,5}$ Daniel M. T\"obbens,$^{1}$ Thomas Unold,$^{1}$ J\'an Min\'ar,$^{2}$ and Oliver Rader$^{1,\ast}$}

\affiliation{$^1$ Helmholtz-Zentrum Berlin f\"ur Materialien und Energie, Albert-Einstein-Str. 15, 12489 Berlin, Germany}
\affiliation{$^2$ New Technologies Research Centre, University of West Bohemia, Z-301 00 Pilsen, Czech Republic}
\affiliation{$^3$ Leibniz Institute for Solid State and Materials Research, IFW Dresden, 01069 Dresden, Germany}
\affiliation{$^4$ Joint Laboratory “Functional Quantum Materials” at BESSY II, 12489 Berlin, Germany}
\affiliation{$^5$ IMDEA Nanoscience, C/ Faraday 9, Campus de Cantoblanco, 28049 Madrid, Spain}
\affiliation{$^\ast$\ Corresponding author.}

\begin{abstract}
\bf 
Halide perovskites are a promising class of materials for optoelectronic and photovoltaic applications, exhibiting high power conversion efficiency due to strong light absorption and long carrier diffusion lengths. While various aspects of their crystal and electronic structure have been studied, we identify a fundamental property previously overlooked that may significantly impact their efficiency. We demonstrate that halide perovskites realize a three-dimensional (3D) Lieb lattice, giving rise to a gapped 3D Dirac cone of spin-1 fermions. This leads to a fivefold reduction in effective mass compared to a conventional cubic structure and suppressed carrier backscattering due to Klein tunneling. Our conclusions are supported by band structure calculations and angle-resolved photoemission spectroscopy from CsPbBr$_3$ and CsSnBr$_3$. In particular, we reveal the transformation of the flat band of the Lieb lattice and the emergence of a dark corridor effect in photoemission from the Dirac cone, which increases as the band gap is decreased from CsPbBr$_3$ to CsSnBr$_3$.
\end{abstract}

\maketitle

\section{Introduction}

In the past decade, metal halide perovskite (MHP) semiconductors have attracted enormous research interest owing to their remarkably high photovoltaic efficiency, long carrier lifetime and diffusion lengths, defect tolerance, tunable band gap, facile synthesis, and low manufacturing costs for photovoltaics, lighting, radiation detection, and single-photon emission \cite{Ahmad22,Park22,Zhu24}. While the precise reasons for their remarkable efficiency have remained unclear, the success of MHPs is expected to originate in their electronic structure. It features band gaps close to the ideal Shockley-Queisser gap and is appropriate for efficient tandem cells with silicon \cite{Ashouri20}. The valence and conduction bands are highly dispersive with relatively small effective masses, facilitating electron and hole transport as the effective mass is inversely proportional to the carrier mobility. The relatively long diffusion length for a typically poorly ordered system of several micrometers \cite{Xing13,Stranks2013} can be explained by the low calculated effective mass \cite{Brivio14} and long charge recombination lifetimes. According to bandstructure calculations, the bands near the extremal points are nonparabolic due to a large spin-orbit splitting, rendering the effective mass momentum dependent \cite{Brivio14}. A low experimental effective mass of $0.1~m_0$ has been deduced from Landau level spectroscopy of MAPbI$_3$ (MA = methylammonium) \cite{MiyataNP15}. Calculations by Frost et al. indicated that the valence band maximum (VBM) is situated at the corner point R of the cubic Brillouin zone (BZ) and formed by antibonding orbitals as a characteristic feature of the MHPs \cite{Frost14}. Goesten and Hoffmann predicted a band that mirrors the dispersion of the upper valence band at high binding energies, which supports the R point as accommodating the most bonding and antibonding pair of orbitals \cite{Goesten18}. The location of the VBM at R and the hole effective mass $m^*$ can be measured by angle-resolved photoemission spectroscopy (ARPES) and was determined for MAPbBr$_3$ as $0.25(8)~m_0$ \cite{ZuJPCL19}. The error bar reflects the limitations in attaining {\bf k}$_\perp$\ (i.e., the electron wave vector perpendicular to the surface) of the VBM when the photon energy is fixed at 21.2~eV \cite{ZuJPCL19}. Hybrid organic-inorganic MHPs are unstable in prolonged ARPES experiments with synchrotron radiation. Still, the more robust CsPbBr$_3$\ emerged as a suitable model system with solar-cell performance similar to hybrid MAPbBr$_3$ \cite{Wang23}. For CsPbBr$_3$, $m^*$ was determined by ARPES at first as $0.26(2)~m_0$ \cite{Puppin20}, but after reassessment of the photon energy needed to access the VBM, it was refined to $0.203(16)~m_0$ \cite{Sajedi22}. Spin-orbit splitting was found to be negligible in ARPES \cite{Sajedi20,Nandi20}.

As a crucial feature distinguishing MHPs from classical tetrahedrally bonded semiconductors, the Pb 6$s^2$ (and Sn 5$s^2$) lone-pair electrons have been identified \cite{Fabini20}.
According to Fabini et al., \cite{Fabini20}, they have a decisive impact on electronic properties, causing high polarizability and the wide valence band and low effective mass of MHPs.
This view is contradicted by the work of Wuttig et al. \cite{Wuttig22}, where the unique properties are attributed to a metavalent bonding mechanism. The effective mass rather increases with the increasing contribution of Pb $s$ states. Instead of the Pb $s^2$ lone pair, the Pb-Br bonding is being held responsible for the low effective mass and the high optical absorption \cite{Wuttig22}.

In the following, we will give an   interpretation  different from the previous analyses \cite{Frost14,Goesten18,Fabini20,Wuttig22,LeeJH16} of the reason for the main features of the band dispersion, its broad width, the gap, and the small effective mass of the valence band. 
We will show, on the one hand, that the crystal structure of halide perovskites is a realization of a 3D Lieb lattice and, on the other hand, how the distinctive electronic properties of the 3D Lieb lattice manifest in the electronic structure of halide perovskites, using CsPbBr$_3$ and CsSnBr$_3$ as examples.

\begin{figure}[ht]
	\centering
	\includegraphics[width=1.0\textwidth]{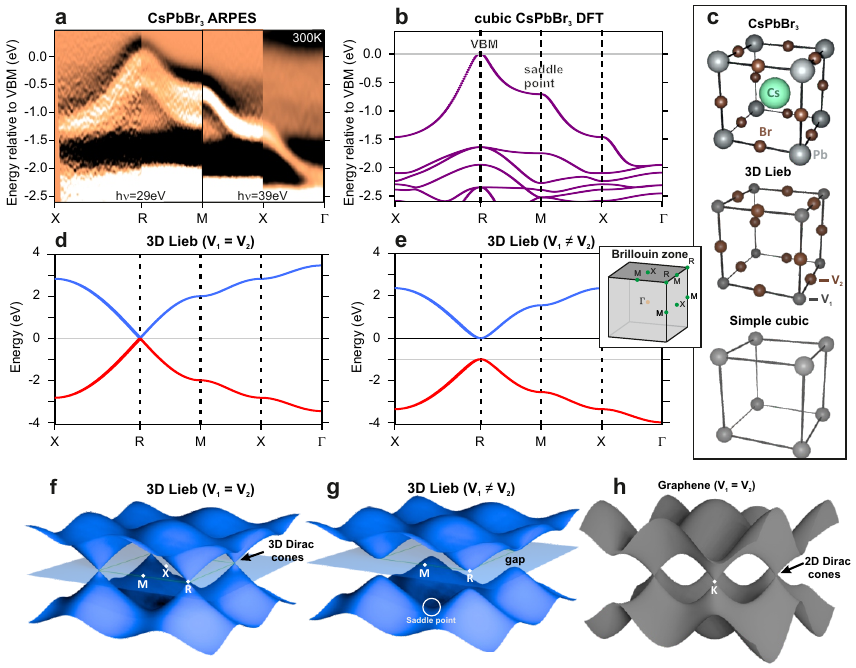}
	\caption{
	\label{DC} 
		{\bf 3D Dirac cone in metal halide perovskites.}
		{\bf a} ARPES intensity (second derivative with respect to energy) and 
		{\bf b} DFT calculation of CsPbBr$_3$.
		{\bf c} Crystal structures of CsPbBr$_3$, 3D Lieb and the simple cubic lattice.
		{\bf d,e} Tight-binding calculations of the 3D Lieb lattice without and with the gap.
		{\bf f,g} The same calculations presented differently, with 3D Dirac cones and saddle points indicated.
		{\bf h} Graphene bandstructure for comparison to 3D Lieb lattice in panel {\bf f}.
		Inset in panel {\bf e} shows the Brillouin zone for the lattices in panel {\bf c}.
	}
\end{figure}

\section{Gapped spin-1 Dirac cone}

{\bf Figure ~\ref{DC} a} summarizes our ARPES band structure measurements in 3D momentum space. Our density functional theory (DFT) band structure calculation for cubic CsPbBr$_3$\ in {\bf Fig.~\ref{DC} b} reproduces the main features of the experiment. (Exceptions are the split-off bands along XR and RM which are replica bands \cite{Park24}). 
{\bf Fig. ~\ref{DC}} illustrates our novel interpretation of the electronic structure of halide perovskites.
First, we note that the cation (here, the Cs) does not contribute to the calculated band structure from the vicinity of the band gap down to 7 eV (see \SNprojections). The remaining atoms form a cubic lattice where Pb atoms occupy the corners, while Br atoms are positioned at the edge centers ({\bf Fig.~\ref{DC} c}). This extends the Lieb lattice \cite{Lieb89} into the third dimension. We calculate the band structure of such a 3D Lieb lattice by a tight binding approach. {\bf Fig.~\ref{DC} d} and {\bf f} show the resulting band structure for the case that atoms at the corners and edge centers are at the same potential. At the R points we see Dirac cones, characteristic of the Lieb lattice \cite{WeeksFranzPRB10}. In the present case, however, we obtain 3D Dirac cones, symmetric in all three wave vector directions (see \SNlieb\ for more details). A gap opens for non-identical atoms (i.e., a potential difference V$_1$ $\ne$ V$_2$ in the model) ({\bf Fig.~\ref{DC} e, g}). In {\bf Fig.~\ref{DC} e}, we see that the resulting dispersion strongly resembles the experimental data  ({\bf Fig.~\ref{DC} a}) and the DFT calculation ({\bf Fig.~\ref{DC} b}). The 3D Lieb lattice is a quadripartite lattice, similar to the bipartite graphene in 2D, but with four sublattices. The graphene bandstructure is presented in {\bf Fig.~\ref{DC} h} for comparison. We note that MHPs are related to 3D Dirac semi-metals in a similar way as hexagonal boron nitride is related to graphene, namely by breaking the sublattice symmetry and resulting gap opening.

\begin{figure}[ht]
	\centering
	\includegraphics[width=1.0\textwidth]{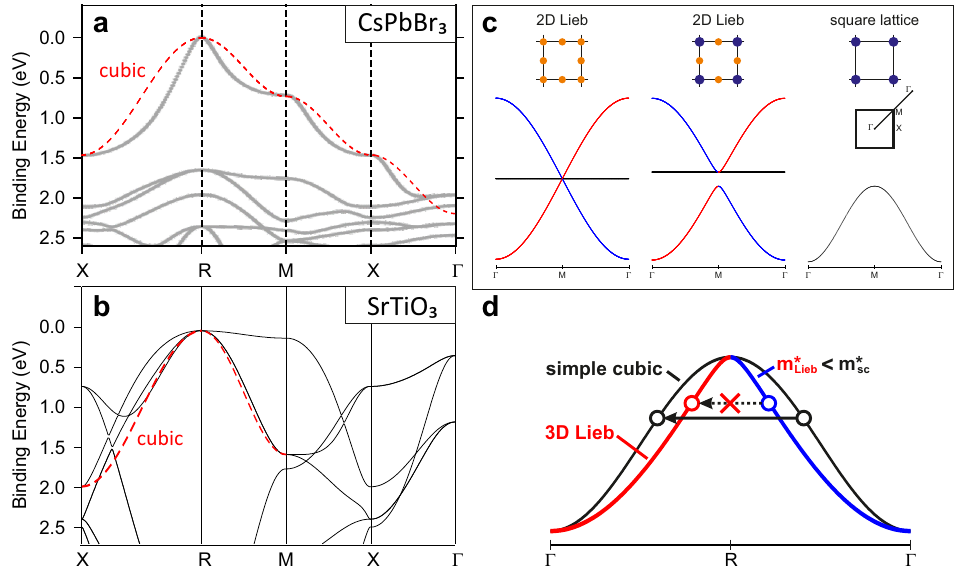}
	\caption{
	\label{LiebMeff} 
		{\bf Effective mass and Klein tunneling.} 
		{\bf a} Comparison of CsPbBr$_3$ and simple cubic lattice bandstructures.
		{\bf b} Comparison of SrTiO$_3$ and simple cubic lattice bandstructures. 
		{\bf c} 2D example of pristine and gapped Lieb lattice bandstructures and of the square lattice bandstructure.
		{\bf d} Comparison of the 3D Lieb lattice and simple cubic lattice band shapes, effective masses and possibility of backscattering.
	}
\end{figure}

\begin{figure}[ht]
	\centering
	\includegraphics[width=0.8\textwidth]{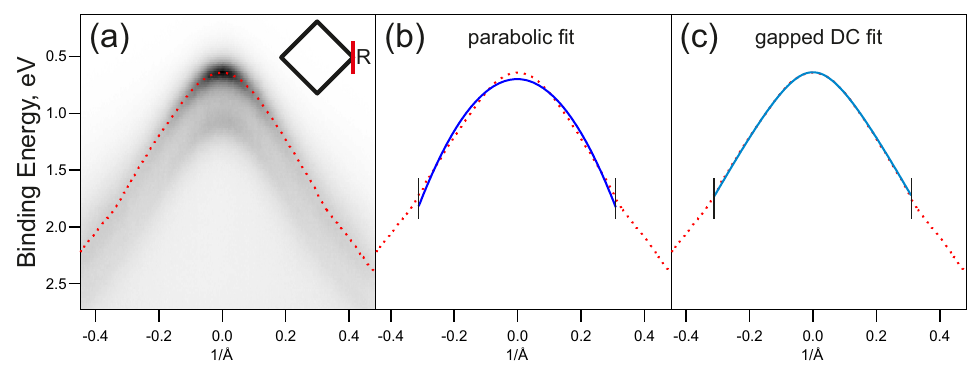}
	\caption[figure]
	{
	\label{fitting} 
	{\bf Gapped Dirac cone dispersion.}
	(a) Experimental data: ARPES of CsSnBr$_3$ with the upper band dispersion marked by dotted line. 
	(b) Parabolic function fit of the dispersion compared to 
	(c) gapped Dirac cone (DC) function fit.
	} 
\end{figure}

Even with the gap, the bands of the 3D Lieb lattice (and CsPbBr$_3$\ correspondingly) have the peculiar dispersion of a Dirac cone with a strongly curved band dispersion around the VBM.
This is illustrated in more detail in {\bf Figures~\ref{LiebMeff} a,b}, which compare the band structure of halide and oxide perovskites with that of a simple cubic lattice with one atom per unit cell. While the latter exhibits a cosine-like dispersion near the band maxima, the gapped Dirac cone in halide perovskites results in a flatter band bottom and a narrower top near the VBM. 
Fitting of the R-point band dispersion with a gapped Dirac cone function provides a significantly better match over a broad wave vector range compared to a parabolic band fit ({\bf Figure~\ref{fitting}} and \SNDCfit).

The effective mass in CsPbBr$_3$\ is therefore about five times lighter (and the mobility five times higher) than in the case where the perovskite or 3D Lieb lattice is replaced by the simple cubic lattice of {\bf Fig.~\ref{DC} c} (see also comparison with other oxides in \SNoxides). MHPs have so far not been considered as Dirac materials \cite{Wehling14} except for a hypothetical topological phase transition, which would pass through an accidental semimetal phase under hydrostatic pressure \cite{Jin12, Yang12}. This pressure-induced topological insulator phase with a protected 2D Dirac cone has not been realized yet, possibly because it would require blocking the octahedral tilts and rotations that MHPs undergo under pressure \cite{Liu16}.

\section{Suppressed backscattering}

The second distinctive property of the Lieb lattice is Klein tunneling \cite{Klein29}, known from graphene as suppressed electron backscattering from an electrostatic potential barrier of arbitrary height and thickness at certain incidence angles  \cite{Katsnelson06,Urban11,Bardarson09,Wehling14}. 
In \SNbackscattering\ we present the model and  the results for the calculated backscattering for both 2D and 3D Lieb lattices. At zero gap, backscattering is fully suppressed; in the presence of a gap, it becomes nonzero but remains suppressed, depending on the gap size. {\bf Fig.~\ref{LiebMeff} c} compares the band structures in the 2D case for simplicity: 2D Lieb with equal atoms, 2D Lieb with sublattice asymmetry, and the square lattice. In {\bf Fig.~\ref{LiebMeff} d} the band structures are overlaid to emphasize the reduced effective mass and scattering possibility. 
Red and blue colors of bands in panels {\bf c} and {\bf d} represent the  sign of the first wave vector component, as discussed in \SNbackscattering, which can be viewed as analogous to the pseudospin in graphene.

When discussing Klein tunneling, it is important to note that the Dirac cone in graphene relates to spin-1/2 fermions. The Dirac-like cone of the 2D Lieb lattice is different, with three instead of two inequivalent lattice sites leading to formally spin-1 instead of spin-1/2 chiral fermions. Interestingly, this higher angular momentum leads to Klein tunneling with even larger transmission for large off-normal angles \cite{ShenPRB10}. In addition, zero backscattering occurs for all angles at energy $V/2$ \cite{ShenPRB10}, termed super Klein tunneling. The effect of the gap is only in reducing the transmission while the Klein tunneling effect persists \cite{Jakubsky23}, see \SNbackscattering. The same holds for a 3D Lieb lattice where only a 2-fold degeneracy of the flat band is added. We expect that Klein and super Klein tunneling of photoexcited carriers directly contribute via the scattering rate to the charge carrier lifetime and thus to the mobility in halide perovskites.

\section{Destructive interference}

\begin{figure}[ht]
	\centering
	\includegraphics[width=1.0\textwidth]{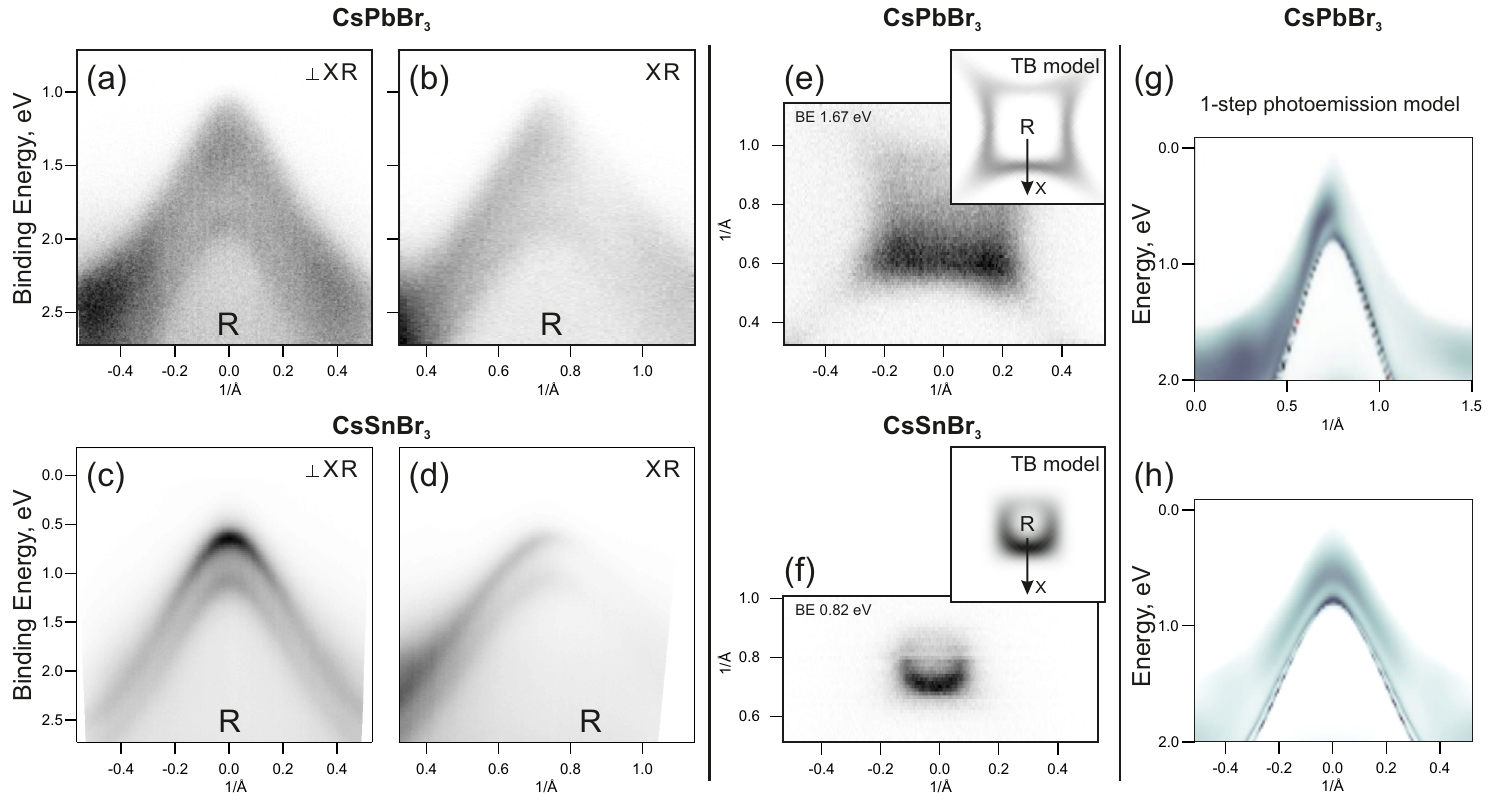}
	\caption{
        \label{FIGasymm} 
	    {\bf Dark corridor effect.} 
		{\bf a,b} ARPES measurement of CsPbBr$_3$ single crystal at h$\nu$=29 eV in $\perp\rm{XR}$ {\bf a} and $\rm{XR}$ {\bf b} directions. 
		{\bf c,d} Similar measurements for a CsSnBr$_3$ film on Au(100) at h$\nu$=27 eV.
		{\bf e,f} Experimental constant energy cuts and model tight-binding (TB) calculations of the intensity distribution at selected binding energies (BE).
		{\bf g,h} One-step photoemission calculation for CsPbBr$_3$ at 29 eV along XR {\bf g} and  perpendicular to XR {\bf h}.
	}
\end{figure}

Considering graphene, another manifestation of the two sublattices, in addition to the Dirac cone and the absence of backscattering, is destructive interference of photoelectrons and suppression of the Dirac cone at certain areas of momentum space \cite{Shirley95,Bostwick07}, the so-called {\it dark corridor} effect \cite{Gierz11}. The effect is connected to the Dirac cone gap size and leads to maximum intensity asymmetry at smaller gaps \cite{Bostwick07}. MHPs have very different dimensionality and lattice symmetry (3D cubic) compared to graphene (2D hexagonal). In view of the Dirac cone origin of the valence band maximum and the presence of several sublattices, we explore the effect in ARPES and tight-binding calculations. 

{\bf Figure~\ref{FIGasymm}} shows ARPES measurements of CsPbBr$_3$ and CsSnBr$_3$ samples in two experimental geometries. Perpendicular to XR, the intensity is high and symmetric about R. Along XR, there is strong suppression of the Dirac cone in the second Brillouin zone. {\bf Figs.~\ref{FIGasymm} e,f} show constant energy cuts with an intensity distribution that resembles the {\it dark corridor} effect in graphene. Our tight-binding calculations presented in insets in {\bf Figs.~\ref{FIGasymm} e,f} support the experimental observation. 1-step photoemission calculations also provide a similar intensity distribution. The measured intensity asymmetry for CsPbBr$_3$ is about 4:1 whereas for CsSnBr$_3$ it is about 8.5:1. Given the smaller band gap of CsSnBr$_3$, this is a strong confirmation of the Lieb lattice origin of the Dirac cone and further support of its associated properties.

\begin{figure}[ht]
	\centering
	\includegraphics[width=1.0\textwidth]{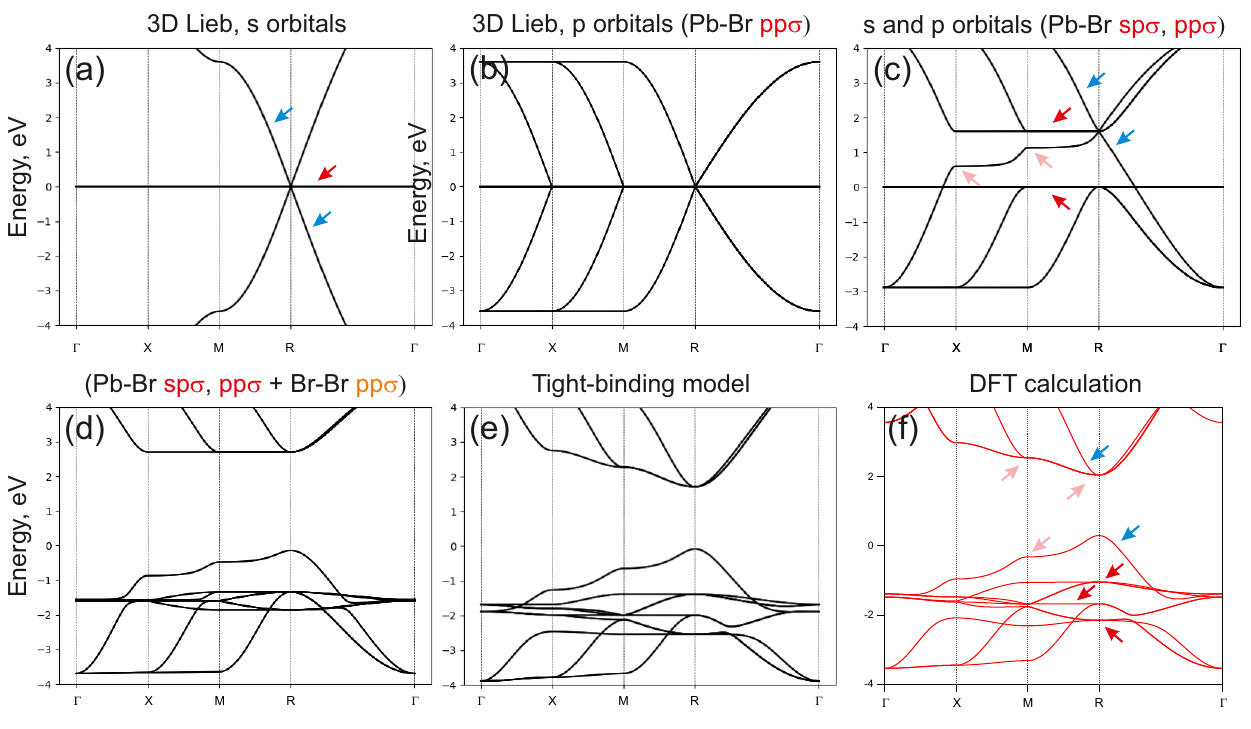}
	\caption{
        \label{Flat} 
		{\bf Evolution of the flat bands.}
		{\bf a} A tight-binding model of the 3D Lieb lattice based on $s$ orbitals only. The red arrow indicates the flat band, and the blue arrows indicate the Dirac cone.
		{\bf b} Same 3D Lieb lattice with $p$ orbitals only and $pp\sigma$ interaction.
		{\bf c} Mix of $s$ and $p$ orbitals. Pb and Br potentials are selected so that the Dirac cone gap stays closed.
		{\bf d} Addition of Br-Br interaction and sublattice potential difference opens the gap.
		{\bf e} The tight-binding model with more interactions taken into account and fitting to the DFT calculation in panel {\bf f}. Blue arrows show the gapped Dirac cone, red arrows indicate flat bands, pink arrows show quasi-flat areas. 
	}
\end{figure}

\section{Flat bands}

Another key feature of the Lieb lattice is the presence of a flat band in its electronic structure which, for the case of the 2D Lieb lattice, is visible in {\bf Fig.~\ref{LiebMeff} c}. Adding a third dimension shifts the Dirac cone to the R point of the Brillouin zone and doubles the flat band ({\bf Fig.~\ref{Flat} a}). However, this classical picture is valid for the model involving only $s$-orbital interactions. When considering the interaction of $p$-orbitals in the 3D Lieb lattice structure, the scenario becomes more complex with nodal planes along all Brillouin zone boundaries ({\bf Fig.~\ref{Flat} b}). To explore this in detail, we included both $s$- and $p$-orbitals, choosing parameters that preserve the Dirac cone while splitting the flat band into two pure flat bands ({\bf Fig. \ref{Flat}c}). Additionally, several quasi-flat regions appear within the Brillouin zone, where the bands are flat locally. Further enhancing the sublattice asymmetry and introducing $\sigma$-type Br-Br interactions yields the band structure in {\bf Fig.~\ref{Flat} d}, which closely resembles the CsPbBr$_3$ band structure but with a fully flat conduction band minimum. Including additional $pp\pi$- and $pd\pi$-type interactions adds some dispersion to the conduction band minimum, resulting in the tight-binding band structure shown in {\bf Fig.~\ref{Flat} e}, which reproduces well the DFT calculation in {\bf Fig.~\ref{Flat} f}. Here we do not take spin-orbit coupling into account for simplicity.

The step-by-step tight-binding construction presented here provides a clear understanding of the origin of the bands observed in DFT ({\bf Fig.~\ref{Flat} f}). The resulting band structure features a gapped Dirac cone, indicated by blue arrows, characterized by a small VBM effective mass and suppressed backscattering, as discussed above. Examining the other bands in more detail, we find that some bands with minimal or negligible energy dispersion in DFT are inherently flat bands, originating from the flat bands of the 3D Lieb lattice.

\

\section{Discussion}

\indent Our finding that CsPbBr$_3$\ features a gapped Dirac cone has, to our knowledge, not been considered so far. It can be generalized to hybrid MHPs since the organic cations do also (like the Cs here) not participate in the electronic structure, and MHPs can generally be represented by 3D Lieb lattices (see \SNhybrid). Thus, the small effective masses of electrons and holes at the band edges are an intrinsic feature of MHPs, arising from the Dirac cone associated with the 3D Lieb lattice. Conversely, MHPs are the first realization of a 3D Lieb lattice in an actual material. Until now, Lieb lattices had only been achieved in 2D through atomic manipulation on Cu surfaces \cite{Drost17,Slot17}. A future task will be quantifying the reduced scattering properties of spin-1 fermions in MHPs due to Klein tunneling.
In graphene, Klein tunneling leads to an anti-localizing effect, enhancing the conductivity; most importantly, backscattering is avoided at the boundaries of charge puddles
\cite{Peres10}. Such influence may be in place in highly defective and/or electrically polarized MHP films potentially explaining why they are successfully competing with single-crystal Si solar cells. For the spin-1/2 fermions in graphene, Klein tunneling has been proven in electrical transport experiments with a gate-induced potential step \cite{Stander09,Young09}. For a 3D system, the scattering properties of Dirac cones may, instead, be probed by quasiparticle interference \cite{Avraham18}. This requires that the system is suitable for scanning tunneling microscopy, which has been demonstrated \cite{Hieulle20}. Moreover, the unconventional transport properties of the Dirac cone can also be probed by magnetotransport, and spin-1 fermions can be distinguished from the conventional spin-1/2 Dirac cone by their different Landau level spectra \cite{Urban11}. 
A further question is whether also phononic transport in MHPs is affected by the Lieb lattice since phononic super Klein tunneling \cite{Jakubsky23} has recently been demonstrated in a macroscopic phononic system \cite{Wu24} and could affect polaron transport in MHPs. 
As a final remark, a MHP could potentially host a topological semimetal with a protected Dirac point, a feature that is unattainable in oxide perovskites \cite{WeeksFranzPRB10}. Chirality, which has recently been realized in MHPs \cite{Long20}, could help protect the Dirac point as demonstrated recently with chiral crystal structures such as CoSi and PtAl, which share the enhanced angular momentum of multifold fermions with the spin-1 Dirac fermions of MHPs \cite{RaoZNature19,SchroterNP19}.

\section{Acknowledgments}

S.A.K. and J.M. thank the QM4ST project financed by the Ministry of Education of the Czech Republic grant no. CZ.02.01.01/00/22\_008/0004572, co-funded by the European Regional Development Fund.

Use of the Helmholtz Innovation Lab HySPRINT for sample preparation is gratefully acknowledged.

\ 

\newpage

\def\StructGrapheneAbs{$7\sqrt{3}\times7\sqrt{3}$}
\def\LDP{LD-phase}
\def\HDP{HD-phase}
\def\AB{{\it A-B}}
\def\Ef{$E_{\rm F}$}
\def\Tc{$T_{\rm C}$}
\def\kpara{{\bf k}$_\parallel$}
\def\kparax{{\bf k}$_{\parallel,x}$}
\def\kparay{{\bf k}$_{\parallel,y}$}
\def\kz{{\bf k}$_\perp$}
\def\kperp{{\bf k}$_\perp$}
\def\Gbar{$\overline{\Gamma}$}
\def\Mbar{$\overline{\rm M}$}
\def\Xbar{$\overline{\rm X}$}
\def\dirGX{$\overline{\rm \Gamma}-\overline{\rm X}$}
\def\dirMXM{${\rm M}{\rm X}{\rm M}$}
\def\dirRMR{${\rm R}{\rm M}{\rm R}$}
\def\dirXRb{${\rm X}{\rm R}$}
\def\dirGMb{${\Gamma\rm M}$}
\def\dirGY{$\overline{\rm \Gamma}\overline{\rm Y}$}
\def\dirGK{$\overline{\rm \Gamma}\overline{\rm K}$}
\def\dirGM{$\overline{\rm \Gamma}\overline{\rm M}$}
\def\dirGS{$\overline{\rm \Gamma}\overline{\rm S}$}
\def\dirGN{$\overline{\rm \Gamma}\overline{\rm N}$}
\def\dirGMtic{$\overline{\rm \Gamma}_{\rm TiC}-\overline{\rm M}_{\rm TiC}$}
\def\dirGKtic{$\overline{\rm \Gamma}_{\rm TiC}-\overline{\rm K}_{\rm TiC}$}
\def\dirMKtic{$\overline{\rm M}_{\rm TiC}-\overline{\rm K}_{\rm TiC}$}
\def\dirGMgr{$\overline{\rm \Gamma}_{\rm Gr}-\overline{\rm M}_{\rm Gr}$}
\def\dirGKgr{$\overline{\rm \Gamma}_{\rm Gr}-\overline{\rm K}_{\rm Gr}$}
\def\dirMKgr{$\overline{\rm M}_{\rm Gr}-\overline{\rm K}_{\rm Gr}$}
\def\dirGMsc{$\overline{\rm \Gamma}_{\rm SC}-\overline{\rm M}_{\rm SC}$}
\def\dirGKsc{$\overline{\rm \Gamma}_{\rm SC}-\overline{\rm K}_{\rm SC}$}
\def\dirMKsc{$\overline{\rm M}_{\rm SC}-\overline{\rm K}_{\rm SC}$}
\def\dirGNhalf{$\frac{1}{2}(\overline{\rm \Gamma}-\overline{\rm N})$}
\def\pntG{$\overline{\rm \Gamma}$}
\def\pntM{$\overline{\rm M}$}
\def\pntK{$\overline{\rm K}$}
\def\pntN{$\overline{\rm N}$}
\def\pntGsc{$\overline{\rm \Gamma}_{\rm SC}$}
\def\pntKsc{$\overline{\rm K}_{SC}$}
\def\pntMsc{$\overline{\rm M}_{SC}$}
\def\pntGtic{$\overline{\rm \Gamma}_{TiC}$}
\def\pntKtic{$\overline{\rm K}_{\rm TiC}$}
\def\pntMtic{$\overline{\rm M}\def\dirGX{$\overline{\rm \Gamma}-\overline{\rm X}$}{\rm TiC}$}
\def\pntGgr{$\overline{\rm \Gamma}_{\rm Gr}$}
\def\pntKgr{$\overline{\rm K}_{\rm Gr}$}
\def\pntMgr{$\overline{\rm M}_{\rm Gr}$}
\def\pntNhalf{ $\overline{\rm N}/2$ }
\def\invA{\AA$^{-1}$}
\def\DCgamma{${\rm DC}_{\overline{\Gamma}}$}
\def\DCNhalf{${\rm DC}_{\overline{\rm N}/{\rm 2}}$}
\def\root33{$\sqrt{3}\times\sqrt{3}$ {\it R}30$^\circ$}
\def\RT3{$\sqrt{3}$}
\def\aR{\alpha_{\rm R}} 
\def\Ga{$\Gamma$}
\def\GM{$\Gamma$M}

\def\twobytwo{$2\times2$}
\def\twotimestwo{$2\times2$}

\def\CPB{CsPbBr$_3$}
\def\CsPbI3{CsPbI$_3$}
\def\CsPbBr3{CsPbBr$_3$}
\def\CsPbCl3{CsPbCl$_3$}
\def\MAPbI3{MAPbI$_3$}
\def\MAPbBr3{MAPbBr$_3$}
\def\MAPbCl3{MAPbCl$_3$}
\def\MAPbX3{MAPb$X_3$}
\def\CsPbX3{CsPb$X_3$}
\def\MAPI{MAPbI$_3$}
\def\MPB{MAPbBr$_3$}
\def\CPB{CsPbBr$_3$}
\def\CsSnBr3{CsSnBr$_3$}

{
\Large
\center
Supplementary Information for
}

{
\Large
\center
Halide Perovskites as Spin-1 Dirac Materials
}

\ 

{\noindent Dmitry Marchenko, Maryam Sajedi, Maxim Krivenkov, Saleem Ayaz Khan, Andrei Varykhalov, Alexander Fedorov, Jaime S\'anchez-Barriga, Daniel M. T\"obbens, Thomas Unold, J\'an Min\'ar, and Oliver Rader}

\newpage

\renewcommand*{\thefootnote}{\fnsymbol{footnote}}
\renewcommand{\figurename}{Fig. S}

\setcounter{figure}{0}

\newlength{\oldparindent}
\newcommand{\noindentsubsection}[1]{
	\setlength{\oldparindent}{\parindent}
	\setlength{\parindent} {0pt}
	\subsection{#1}
	\setlength{\parindent}{\oldparindent}
}

\section*{Supplementary Note 1}
\subsection*{Atomic-orbital-projected DFT}

\begin{figure}[H]
	\centering
	\includegraphics[width=0.9\textwidth]{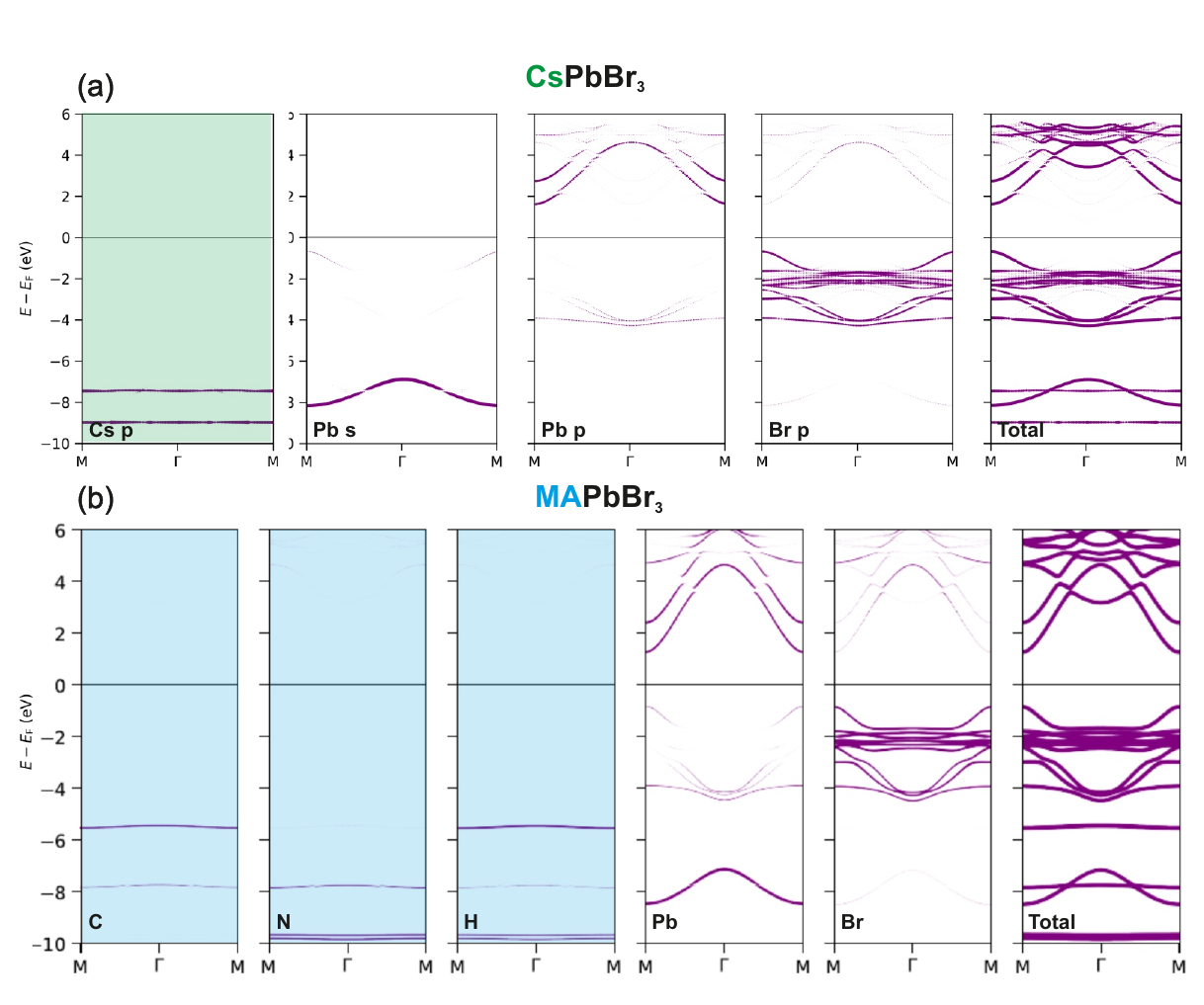}
	\caption[] {\label{orbital} Atomic-orbital-projected DFT band structure for \CsPbBr3 (a) and for MAPbBr$_3$ (b).}
\end{figure}
The atom-orbital-projected band structures for Pb $s$ and Pb, Cs, and Br $p$ orbitals are presented in Figs.~S~1 {\bf a,b}. 
From the orbital-projected DFT calculations it is evident that the Cs spectral weight does not have a significant contribution throughout the dispersive valence and conduction bands. Together with similar results for $d$ orbitals (not shown here), this indicates that Cs atoms do not contribute directly to the bands of interest.
The same is valid for MAPbBr$_3$ (Fig.~S~1 {\bf b}) where the molecular orbitals of the methylammonium cation do not contribute to valence and conduction bands in the region around the Fermi level. 

\newpage

\section*{Supplementary Note 2}
\subsection*{Analytical model for 3D Lieb lattice}

\def\EA{V_{Pb}}
\def\EB{V_{Br}}
\def\cosx{\cos(k_x \frac{d}{2})}
\def\cosy{\cos(k_y \frac{d}{2})}
\def\cosz{\cos(k_z \frac{d}{2})}

\begin{figure}[H]
\centering
\includegraphics[width=0.98\textwidth]{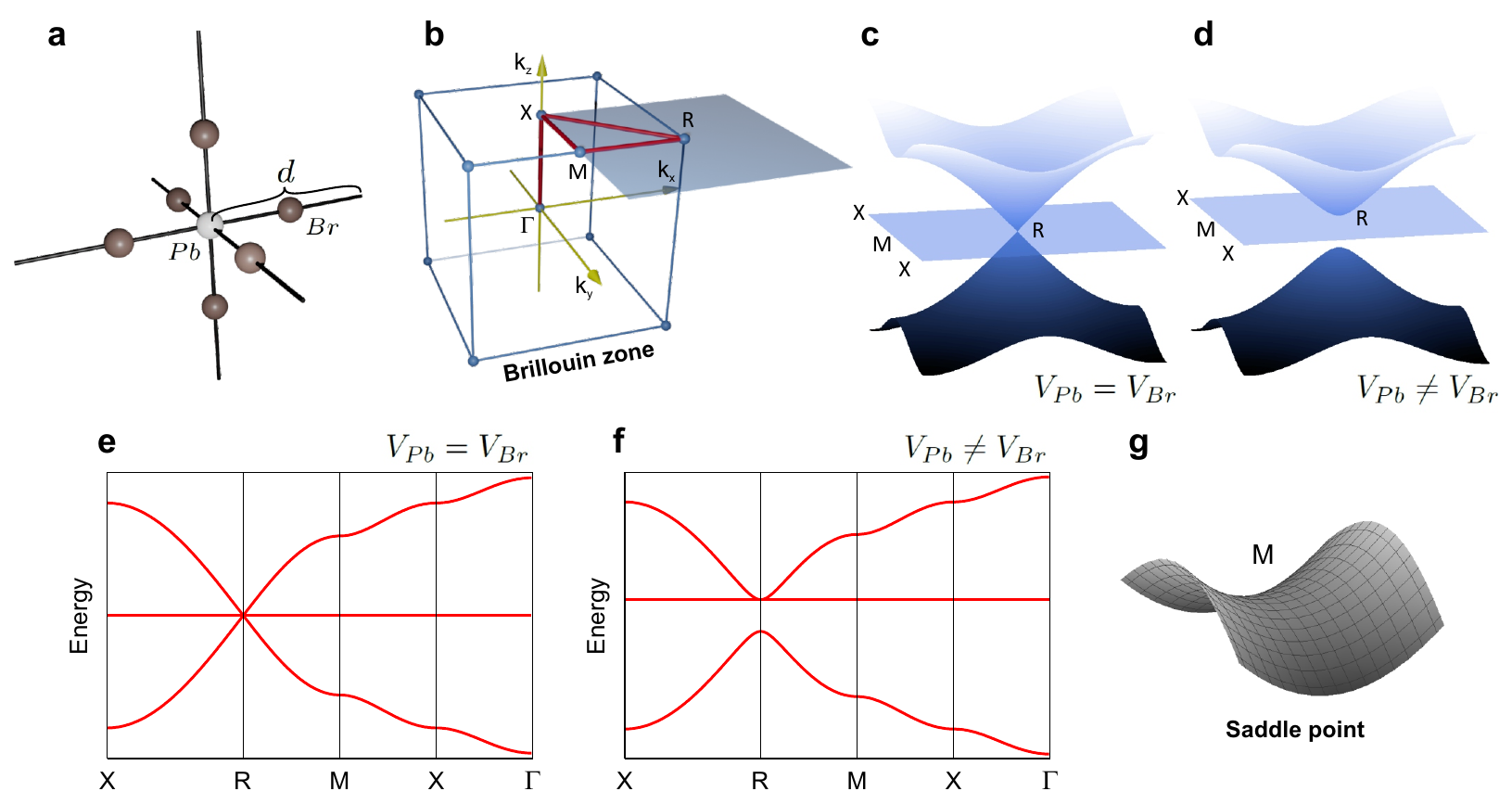}
\caption{
{\bf 3D Dirac cone and saddle point.} 
(a) Crystal structure. 
(b) Brillouin zone. 
(c) 3D Dirac cone at Brillouin zone corner (R point). For better overview, the bands are presented for a square area around R with corners at X points. 
(d) The Dirac cone is gapped when Pb and Br atom potentials are different.
(e) and (f) Bandstructure along XRMX$\Gamma$\ for the intact and gapped cases. 
(g) Saddle point at M.
}
\end{figure}

Figure~S~2 {\bf a} shows the crystal structure of the 3D Lieb lattice as a cubic lattice with the lattice parameter $d$. Pb atoms are located at corners and Br atoms at edge centers at distance of $d/2$ from the Pb atom.  

In the $s$-orbital approximation, the Pb atom interaction with two opposite neighbouring Br atoms along the x-axis can be considered as $f_x(k_x)=t\ \exp(i\ k_x\frac{d}{2}) + t\ \exp(-i\ k_x\frac{d}{2})=2t\ \cosx$, where $t$ is the Pb-Br hopping.
At the Brillouin zone boundary $k_x = \pi/d$ the influences of two opposite Br atoms compensate each other and $f_x$ becomes zero. At the Brillouin zone corner (R-point) the same happens for all three $k$ directions. 

To study the dispersion in detail we can write the Hamiltonian for the 3D Lieb lattice, which has four inequivalent lattice sites, in the following way:

\ 

\renewcommand{\arraystretch}{2}
$\mathcal{H} = 
\begin{pmatrix} 
\EA & f_x(k_x) & f_y(k_y) & f_z(k_z) \\
f_x(k_x) & \EB & 0 & 0 \\
f_y(k_y) & 0 & \EB & 0 \\
f_z(k_z) & 0 & 0 & \EB \\
\end{pmatrix} 
=
\begin{pmatrix} 
\EA & 2t\ \cosx & 2t\ \cosy & 2t\ \cosz \\
2t\ \cosx & \EB & 0 & 0 \\
2t\ \cosy & 0 & \EB & 0 \\
2t\ \cosz & 0 & 0 & \EB \\
\end{pmatrix}  
$

\ 

where $\EA$ and $\EB$ are potentials of Pb and Br atoms, respectively. 

\ 

The solution is:

$\varepsilon_1 = \varepsilon_2 = \EB$
~~~~
$\varepsilon_{3,4} = \frac{1}{2} \left( (\EA+\EB) \pm \sqrt{(\EA-\EB)^2+16t^2\ [\cosx^2+\cosy^2+\cosz^2]} \right)$

\

For $\EA = \EB$ this is a 3D Dirac cone and a doubly degenerate flat band crossing the Dirac point. For $\EA \neq \EB$ the Dirac cone is gapped. Corresponding pictures around the R point are presented in  Figs.~S~2 {\bf c},{\bf d}. Dispersions along XRMX$\Gamma$\ path are presented in Figs.~S~2 {\bf e},{\bf f}. Detailed view around the M point in Fig.~S~2 {\bf g} demonstrates a saddle point.

\

\

\

\

\newpage

\section*{Supplementary Note 3}
\subsection*{Gapped Dirac cone fit function}

A parabolic fit can always be performed locally at the apex of a curve, which is often used to determine the effective mass at a band maximum. However, the result can be highly sensitive to the wave vector range chosen for the fit, potentially overlooking important information about the overall band dispersion.

Here, we extend the analysis to a wider wave vector range to examine the band dispersion shape. Using ARPES data from \CsSnBr3, we show that fitting with a gapped Dirac cone provides a significantly better match over a broad wave vector range compared to a parabolic band fit.
Figure 3 of the main text presents the ARPES data along with an automatic determination of the band dispersion, obtained by fitting each k-slice with two Gaussian curves and a background. The extracted dispersion is then compared to a parabolic function in panel {\bf b} and to a gapped Dirac cone function in panel {\bf c}.

The gapped Dirac cone function was selected in the form
$E(k) = A\sqrt{ \Delta^2 + t^2k^2 } + E_0$, 
where $\Delta$ denotes the relative bandgap contribution, $t$ and $A$ - wave vector and energy scaling, and $E_0$ - energy shift. Based on the fitting, the gap can be estimated as $2A\Delta = 1.3$ eV. 

\ 

\newpage

\section*{Supplementary Note 4}
\subsection*{Effective mass in MHPs and oxide perovskites}

The effective mass in \CsPbBr3\ is small ($m^*\sim~0.20~m_0$) compared to the basic cosine-like dispersion relation, which appears if we replace the 3D Lieb lattice by a simple cubic lattice (m*= $0.65~m_0$).

Table I shows for a few representative oxide perovskites that hole masses are typically very high, which holds for oxides in general \cite{Hautier13}. 

\begin{table}[http]
	\centering
	\begin{tabular}{|c|c|}
		\hline
		{\bf Oxide perovskite} & {\bf Hole effective mass ($m^{*}_h$/$m_0$)} \\ 
		\hline 
		SrTiO$_3$  & 2.8 to 3.5 at \Ga\ \cite{Wunderlich2009} \\
		\hline
		LaFeO$_3$ & 2.11 to 3.28 at \Ga\ \cite{Singh18} \\
		\hline
		BaTiO$_3$ & 1.1 to 8 at A \cite{Bagayoko1998}  \\
		\hline
	\end{tabular}
	\caption[]{Hole effective masses calculated by DFT for oxide perovskites in the literature. The spread of the values is due to an anisotropy of the effective mass. The DFT band gaps (2~eV for SrTiO$_3$ and LaFeO$_3$ and 2.6~eV for BaTiO$_3$) are comparable to those of \CsPbBr3. The values for BaTiO$_3$ relate to the tetragonal structure.}
	\label{oxides} 
\end{table}

\

\newpage

\section*{Supplementary Note 5}
\subsection*{Suppressed backscattering}

\def\cosk{\cos(k \frac{d}{2})}

\begin{figure}[H]
\centering
\includegraphics[width=0.98\textwidth]{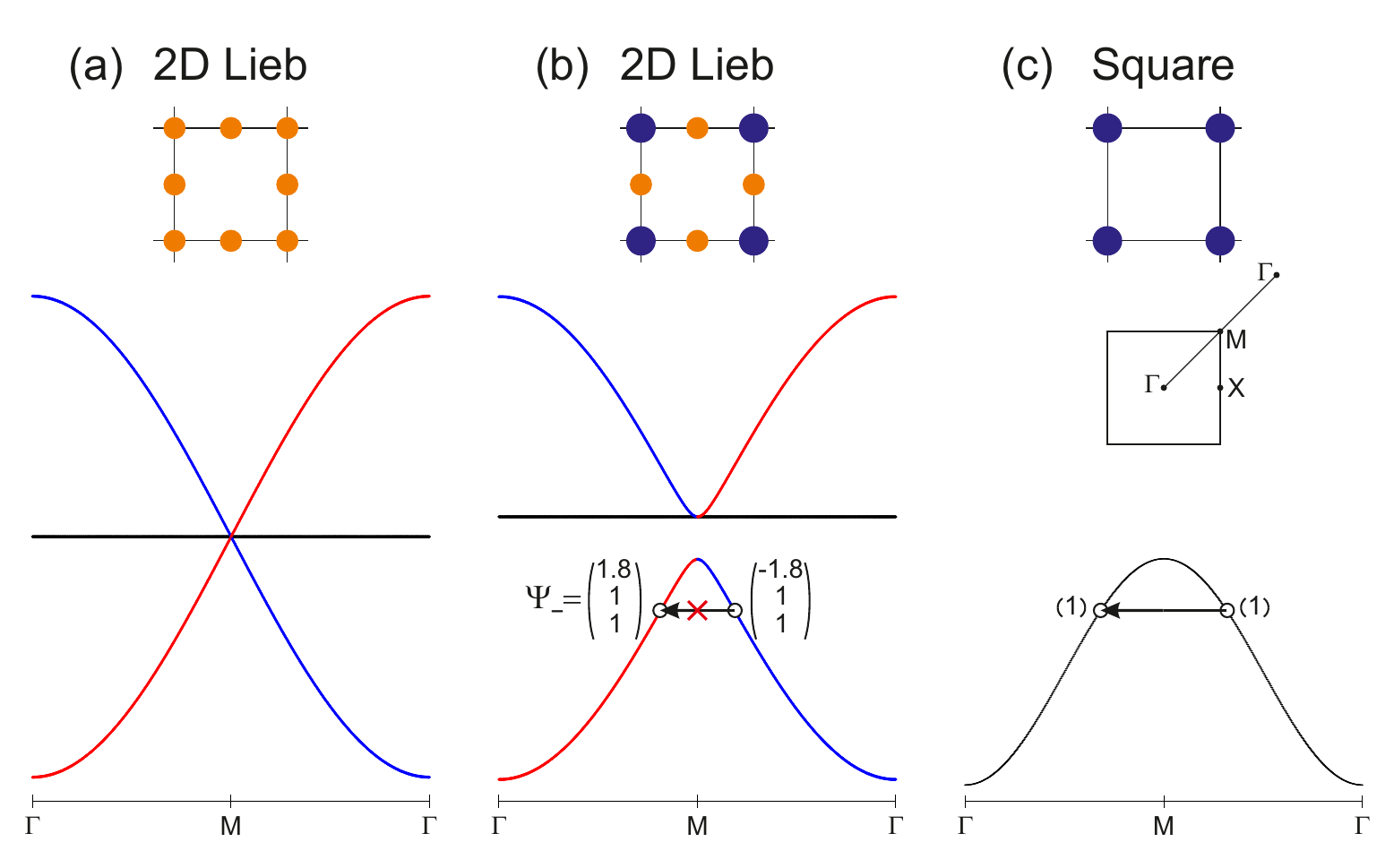}
\caption{
{\bf Comparison of 2D Lieb and square lattices.} 
(a) Electronic structure of 2D Lieb lattice without sublattice asymmetry. 
(b) Gap opening in the case of sublattice asymmetry. 
(c) Electronic structure of square lattice.
Red and blue colors in (a) and (b) represent the sign of the first wave vector component.
}
\end{figure}

For simplicity we demonstrate the mechanism in 2D and for the $\Gamma \rm M \Gamma$ (diagonal) direction in the Brillouin zone, i.e. $k_x = k_y = k$.

\ 
	
${\cal H}=\left( \begin{array}{ccc} 
V_{Pb} & 2t \cosk & 2t \cosk \\ 
2t \cosk & V_{Br} & 0 \\ 
2t \cosk & 0 & V_{Br} \\ 
\end{array}
\right)$	

\ 

We introduce
\ 
$\Delta = V_{Pb} - V_{Br}$
\ 
and
\ 
$\eta_{\pm} = \frac{\Delta \pm \sqrt{\Delta^2 + 32 t^2 \cosk^2}}{4 t \cosk}$

\ 

and write the solution in the general form: 

\ 

$\varepsilon_0 = V_{Br}$
\ \ \ \ \ 
$\varepsilon_{\pm} = \frac{1}{2}\left( V_{Pb}+V_{Br} \pm \sqrt{\Delta^2 + 32 t^2 \cosk^2} \right)$

\ 

\def\arraystretch{1.0}
$\Psi_0 = \left( \begin{array}{c} 
0 \\ 
-1 \\ 
1 \\ 
\end{array}
\right)$
\ \ \ \ \ 
$\Psi_{\pm} = \left( \begin{array}{c} 
\eta_{\pm} \\ 
1 \\ 
1 \\ 
\end{array}
\right)$

\ 

where wave vectors $\Psi_0$, $\Psi_{+}$ and $\Psi_{-}$ are represented as weights on sublattices, without normalization. $+$ corresponds to the upper band, $-$ to the lower one. 

This solution is shown in {\bf Figs.~S~3} {\bf a},{\bf b} for the cases $\Delta = 0$ and $\Delta \neq 0$. By red and blue color we mark the sign of the first wave vector component $\eta_{\pm}$ and see that it changes sign at the M point. At the same time the two other components remain unchanged. Thus the wave vectors before and after the VBM at the M point are not compatible due to their different chirality. This is similar to the sublattice contributions to the phase rotation in graphene around the K point, however, in the Lieb lattice $\eta_{\pm}$ is a real number and not a complex phase. {\bf Figs.~S~3} {\bf c} shows a square lattice band structure with (i) higher effective mass and  (ii) absence of a sublattice effect. 

\begin{figure}[H] 
\centering
\includegraphics[width=0.75\textwidth]{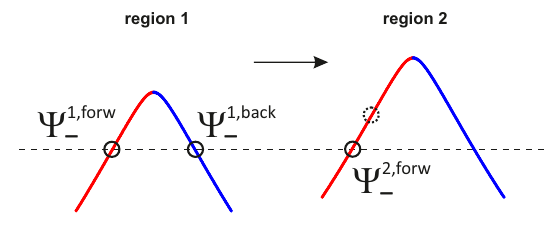}
\caption{
{\bf Suppressed backscattering.} 
Schematic for a thought experiment with potential step encounter. The 2D Lieb lattice is present in region 1 and region 2, but there is an extra potential in region 2 which models the presence of a defect. 
}
\end{figure}

The sublattice effect leads to suppressed backscattering for both electron and hole bands. This can be demonstrated for a simplified example of an electron encountering a region with a different potential (e.g. in the vicinity of a defect) and reflection from the boundary ({\bf Fig.~S~4}). The incident, reflected and transmitted wave functions can be joined in the following way:

\ 

\def\arraystretch{1.0}
$I\ exp(i k_1 x)$
$\left( \begin{array}{c} 
\eta_{-}^{1,forw} \\ 
1 \\ 
1 \\ 
\end{array}
\right)$
$+ R\ exp(- i k_2 x)$
$\left( \begin{array}{c} 
\eta_{-}^{1,back} \\ 
1 \\ 
1 \\ 
\end{array}
\right)$
$= T\ exp(i k_3 x)$
$\left( \begin{array}{c} 
\eta_{-}^{2,forw} \\ 
1 \\ 
1 \\ 
\end{array}
\right)$

\ 

If we consider $x=0$ at the boundary between the regions and $\eta_{-}^{1,back} = - \eta_{-}^{1,forw}$ then we get two conditions:

$I + R = T$ \ \ \ and\ \ \ \ $I\ \eta_{-}^{1,forw} - R\ \eta_{-}^{1,forw} = T\ \eta_{-}^{2,forw}$

\ 

and the solution:

\ 

$R = I\ \frac{\eta_{-}^{1,forw}-\eta_{-}^{2,forw}}{\eta_{-}^{1,forw}+\eta_{-}^{2,forw}}$

\ 

In the limit of small potential distortion, $\eta_{-}^{2,forw}$ is close to $\eta_{-}^{1,forw}$ and the reflection $R$ is close to zero. If $\Delta = 0$ we have $\eta_{\pm} = \pm \sqrt{2}\ {\rm sign}(\cosk)$, $\eta_{-}^{1,forw}=\eta_{-}^{2,forw}=-\sqrt{2}$ and $R = 0$ independently of the size of the potential step.

\ 

The result for the 3D Lieb lattice is similar with 
$\eta^{3D}_{\pm} = \frac{\Delta \pm \sqrt{\Delta^2 + 48 t^2 \cosk^2}}{4 t \cosk}$

\ 

Finally, we obtain zero backscattering for the ungapped case and suppressed but nonzero backscattering in the case of a gap, like in the real perovskite material.

\newpage

\section*{Supplementary Note 6}

\subsection*{Comparison to hybrid organic-inorganic perovskite}

The valence band structure of MAPbBr$_3$ has been calculated ({\bf Figure S~5} {\bf a}). Direct comparison to cubic \CsPbBr3\ (panel {\bf b}) shows that the description as gapped 3D spin-1 Dirac material derived from the 3D Lieb lattice applies as well. The orbital decomposition (see {\bf Fig. S~1} {\bf b}) confirms that the 
molecular orbitals do not contribute in the relevant energy range of the Dirac cone.

\begin{figure}[H]
	\centering
	\includegraphics[width=0.8\textwidth]{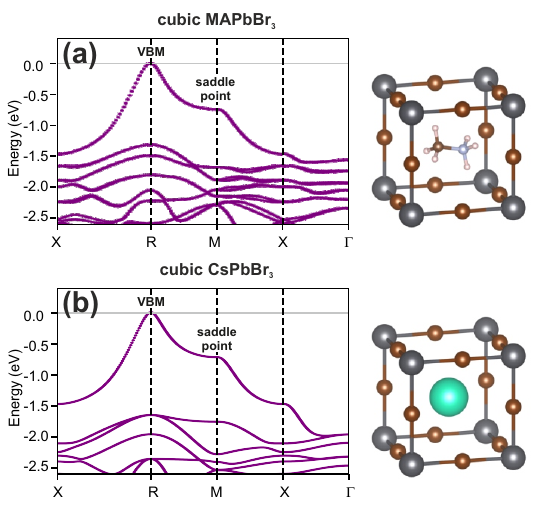}
	\caption[] {\label{cubicdft} Calculated DFT bandstructure for cubic MAPbBr$_3$ and the corresponding bandstructure for cubic \CsPbBr3\ for comparison.}
\end{figure}

\newpage

\section*{Supplementary Note 7}
\subsection*{Sample preparation and experimental details}

Our \CsPbBr3 single crystals have been homegrown by antisolvent vapor-assisted crystallization method proposed by   Rakita \textit{et al.} \cite{RakitaCG16} with some modifications \cite{Sajedi20}. The synthesized samples were attached to molybdenum sample holders using a silver epoxy followed by curing at 80$^{\circ}$C in an Ar atmosphere. Experiments have been done with freshly cleaved surfaces as described previously (see Ref. \cite{Sajedi20} and its Supplemental Material for details on experimental techniques).  

For deposition of thin films of CsSnBr$_3$ on Au(100) we adapted the recipe described in Ref.  \cite{Rieger23}. Using separate home-built evaporators for each material, CsBr and SnBr$_2$ were co-evaporated on the Au(100) substrate which was kept at 120$^{\circ}$C during the deposition. Deposition rates were calibrated with a quartz microbalance to be equal and to provide in total approximately 1 monolayer (ML) of film per minute. Source materials were obtained from Fischer Scientific GmbH.

ARPES has been performed using hemispherical electron energy analyzers and synchrotron radiation at the ARPES-1$^2$ \cite{Varykhalov_1squar18} (Scienta R8000) and Spin-ARPES (Scienta R4000) instrument at BESSY II using linearly polarized undulator radiation. The base pressure of the instruments was better than $2\times10^{-10}$ mbar and the spot size in ARPES $\sim100$ $\mu$m.

\newpage

\section*{Supplementary Note 8}
\subsection*{Theoretical methods}

\indent For band structure calculations the fully relativistic density functional theory (DFT) was used as implemented in the Vienna ab initio Simulation Package (VASP) \cite{Kresse96}. The exchange-correlation functional is approximated by employing the generalized gradient approximation of Perdew-Burke-Ernzerhof (PBE) \cite{Perdew96}, with a full treatment of spin orbit coupling. It should be noted that usually incorporation of spin orbit coupling with PBE underestimates the electronic band gap, mostly due to the splitting of the conduction band minimum \cite{Gill21}. The core and the valence electrons were treated within the projector augmented wave (PAW) method \cite{Blochl94,Kresse99}. 
The lattice parameters were taken from the experimental X-ray diffraction (XRD) patterns.
Tight-binding calculations were conducted with own code and with the use of PySKTB Python package.

\

\phantom{xxxx}

\end{document}